\documentclass[a4paper, amsfonts, amssymb, amsmath, reprint, showkeys, nofootinbib, twoside]{revtex4-1}
\usepackage[english]{babel}
\usepackage[utf8]{inputenc}
\usepackage[colorinlistoftodos, color=green!40, prependcaption]{todonotes}
\usepackage{amsthm}
\usepackage{mathtools}
\usepackage{physics}
\usepackage{xcolor}
\usepackage{graphicx}
\usepackage[left=20mm,right=15mm,top=33mm,columnsep=15pt, bottom=20mm]{geometry} 
\usepackage{adjustbox}
\usepackage{placeins}
\usepackage[T1]{fontenc}
\usepackage{lipsum}
\usepackage{csquotes}
\usepackage{xfrac}

\newcommand{\inprod}[2]{\left(#1\,\middle\vert\,#2\right)}

\usepackage[pdftex, pdftitle={Article}, pdfauthor={Author}]{hyperref} 

\begin{document}
\title{Enhancing gravitational-wave burst detection confidence in expanded detector networks with the BayesWave pipeline}

\author{Yi Shuen C. Lee}
    \email[]{ylee9@student.unimelb.edu.au}
    \affiliation{School of Physics, The University of Melbourne, Victoria 3010, Australia}
\author{Margaret Millhouse}
    \email[]{meg.millhouse@unimelb.edu.au}
    \affiliation{OzGrav, The University of Melbourne, Victoria 3010, Australia}
\author{Andrew Melatos}
    \email[]{amelatos@unimelb.edu.au}
    \affiliation{OzGrav, The University of Melbourne, Victoria 3010, Australia}


\begin{abstract}
The global gravitational-wave detector network achieves higher detection rates, better parameter estimates, and more accurate sky localisation, as the number of detectors, $\mathcal{I}$ increases.
This paper quantifies network performance as a function of $\mathcal{I}$ for \textit{BayesWave}, a source-agnostic, wavelet-based, Bayesian algorithm which distinguishes between true astrophysical signals and instrumental glitches. Detection confidence is quantified using the signal-to-glitch Bayes factor, $\mathcal{B}_{\mathcal{S},\mathcal{G}}$. An analytic scaling is derived for $\mathcal{B}_{\mathcal{S},\mathcal{G}}$ versus $\mathcal{I}$, the number of wavelets, and the network signal-to-noise ratio, SNR$_\text{net}$, which is confirmed empirically via injections into detector noise of the Hanford-Livingston (HL),  Hanford-Livingston-Virgo (HLV), and Hanford-Livingston-KAGRA-Virgo (HLKV) networks at projected sensitivities for the fourth observing run (O4). The empirical and analytic scalings are consistent; $\mathcal{B}_{\mathcal{S},\mathcal{G}}$ increases with $\mathcal{I}$. The accuracy of waveform reconstruction is quantified using the overlap between injected and recovered waveform, $\mathcal{O}_\text{net}$. The HLV and HLKV network recovers $87\%$ and $86\%$ of the injected waveforms with $\mathcal{O}_\text{net}>0.8$ respectively, compared to $81\%$ with the HL network. The accuracy of \textit{BayesWave} sky localisation is $\approx 10$ times better for the HLV network than the HL network, as measured by the search area, $\mathcal{A}$, and the sky areas contained within $50\%$ and $90\%$ confidence intervals. Marginal improvement in sky localisation is also observed with the addition of KAGRA.

\end{abstract}


\maketitle

\section{Introduction} \label{sec:introduction}

The Laser Interferometer Gravitational-Wave Observatory (LIGO) \cite{LIGO2015, 2009RPPh...72g6901A, Harry_2010} has completed three observing runs, O1 \cite{2017PhRvD..95d2003A,2019PhRvX...9c1040A}, O2 \cite{2019PhRvD.100b4017A,2019PhRvX...9c1040A} and O3 \cite{2020NatRP...2..222G} between 2015 to 2020, including joint searches with Italian partner, Virgo \cite{2015CQGra..32b4001A}, in the final month of O2 and the whole of O3. In April 2019, Advanced LIGO commenced its third observing run in collaboration with Advanced Virgo as a three-detector network: the Hanford-Livingston-Virgo (HLV) network. The Kamioka Gravitational Wave Detector (KAGRA) \cite{2019CQGra..36p5008A, 2020arXiv200505574A, 2013PhRvD..88d3007A} also began observing in  February 2020 \cite{2020NatRP...2..222G}.

With access to these upgraded instruments, there is a burgeoning interest in detecting short-duration gravitational wave (GW) signals by combining data from multi-detector networks. These signals typically have durations of milliseconds up to a few seconds, with the most common sources being compact binary coalescences (CBCs) such as black hole or neutron star mergers, along with other potential sources like core-collapse supernovae (SNe) of massive stars \cite{2011LRR....14....1F}, pulsar glitches of astrophysical origin \cite{2014ApJ...787..114S} and cusps in cosmic strings \cite{2005PhRvD..71f3510D}. In addition to these known sources, it is also plausible to detect transient signals of unknown astrophysical origin. 

Searches for generic GW transients, or burst searches, require the ability to distinguish such signals from any noise artefacts present in the detector data. Hence, it is crucial to understand the noise properties of the detector data. Results from the initial LIGO-Virgo science runs revealed non-stationary and non-Gaussian detector noise, which includes short-duration noise transients denoted by the term `glitches' \cite{2008CQGra..25r4004B, 2015CQGra..32k5012A,2012CQGra..29o5002A}. If not accounted for properly, these features could resemble GWs and consequently limit the ability to detect low-amplitude signals. 

Since CBC signals come from known and well-studied sources, such signals are accurately modelled in most regions of parameter space and therefore can be detected with high confidence using matched-filter searches \cite{2016CQGra..33u5004U, 2019arXiv190108580S, 2012PhRvD..86b4012H}. Other GW bursts signals, on the other hand, may originate from either complex or unanticipated sources. Given the stochastic nature and complexity of the potential sources (e.g.~core collapse supernovae), there are no robust models available to date to assist with the searches of generic burst signals, making it challenging to distinguish them from other non-Gaussian features like glitches in the detector data, as well as to accurately reconstruct the underlying signal waveform. 

There are a number of unmodelled burst searches performed in LIGO and Virgo data~\cite{2017PhRvD..95j4046L,cWB}. 
In this work we look at an unmodelled search algorithm called \textit{BayesWave} \cite{2015PhRvD..91h4034L, 2015CQGra..32m5012C, BWIII}, which was proposed to enable the joint detection and characterisation of GW bursts and instrumental glitches. \textit{BayesWave} reconstructs both signals and glitches as a sum of sine-Gaussian wavelets, where the number of wavelets and their parameters are determined via a reversible-jump Markov chain Monte Carlo (RJMCMC) algorithm. Bayesian model selection is then used to determine the likelihood of an event being a true signal, or a noise artefact.

Previous studies have quantified the performance of \textit{BayesWave} in recovering simulated waveforms from simulated noise with a two-detector network (HL network) \cite{2016PhRvD..94d4050L,2016PhRvD..93b2002K}. However, with Virgo joining GW searches alongside the HL-network in O2 and O3, KAGRA coming online towards the end of O3, and future detectors like LIGO-India in the planning stages \cite{2013IJMPD..2241010U}, the network of GW detectors is expanding rapidly. Expanding detector networks will increase the likelihood of detecting more events with higher confidence. These improvements are evident in previous studies and will be elaborated further in Section \ref{sec:network}.

In this paper, we aim to evaluate \textit{BayesWave}'s performance in searching for GW bursts from detector data beyond the HL-network. We achieve this by using \textit{BayesWave} to recover injected signals from simulated noise with the HLV and the HLKV detector networks, and comparing the outcomes with those of the HL network. We quantify the performance of\textit{ BayesWave} based on the following metric: (i) Bayes factor between signal and glitch models, (ii) overlap (match) between injected and recovered waveforms, and (iii) accuracy of recovered sky location. In Section \ref{sec:bw}, we provide a detailed overview of the \textit{BayesWave} algorithm. We derive the analytic scaling relation of the signal-to-glitch Bayes factor in Section~\ref{sec:BF_scaling}. We then discuss the methods of injecting simulated waveforms into simulated detector noise samples in Section \ref{sec:method}, followed by comparisons and analyses of the metrics mentioned above: Bayes Factor in Section \ref{sec:BayesFactor}, overlap in Section \ref{sec:overlap} and sky localisation in Section \ref{sec:skyloc}. Finally, we present a summary of the results along with their implications in Section \ref{sec:summary}.

\section{Benefits of Expanding Detector Networks} \label{sec:network}

Increasing the number of operational ground-based detectors has several major benefits for GW astronomy, including a higher rate of detection of GW transients, and better characterisation of those signals.  Here we discuss some of the benefits of adding new detectors to the existing network.

\subsection{SNR and search volume}
One major advantage of a larger detector network is the ability to confidently detect quieter events. 
The strain amplitude, $s^{(i)}$ in detector $i$ of the network consists of a signal, $h^{(i)}$ (if present) and detector noise, $n^{(i)}$ which can be expressed as
 \begin{equation}
     s^{(i)} = h^{(i)} + n^{(i)}.
 \end{equation}
 The squared matched-filter signal-to-noise ratio (SNR) of signal $h^{(i)}_s$ in detector $i$ is then given by \cite{1994PhRvD..49.2658C}
 \begin{equation}
     \text{SNR}_i^2 = \inprod{h^{(i)}_s}{h^{(i)}_s}
\end{equation}
where $(.|.)$ on the right-hand-side of the expression is the noise-weighted inner product. We define the noise-weighted inner product between two arbitrary waveforms $h_a(t)$ and $h_b(t)$ as \cite{keppel...2009}
\begin{equation}
\inprod{h_a}{h_b} = \int_0^\infty \frac{\tilde{h_a}^*(f)\tilde{h_b}(f) + \tilde{h_a}(f)\tilde{h_b}^*(f)}{S_n(f)} \, df.
\label{eq:noise_weighted_inprod}
\end{equation}
where $\tilde{h}(f)$ is the Fourier-transformed waveform, $\tilde{h}^*(f)$ is its complex conjugate and $S_n(f)$ is the one-sided power spectral density (PSD) of stationary, Gaussian detector noise. 

For a network with $\mathcal{I}$ detectors, the overall network SNR is given by \cite{2016PhRvD..94d4050L}
\begin{equation}
\text{SNR}_\text{net}^2 = \sum_{i=1}^\mathcal{I}\text{SNR}_i^2
\label{eq:snr_net}
\end{equation}
According to Equation \ref{eq:snr_net}, adding more detectors to the network increases the SNR of all detected GW signals. This enables detection pipelines to estimate waveform parameters more accurately \cite{2017CQGra..34q4003G}. With improved parameter estimates, more accurate models can be constructed to represent the detected waveform \cite{2017ApJ...839...15B}.
 
 In addition, the SNR of GW signals scales with luminosity distance, $D_L$ as \cite{2014arXiv1409.0522C}
 \begin{equation}
     \text{SNR}_i \propto \frac{1}{D_L}.
     \label{eq:snr_scale}
 \end{equation}
By combining Equations \ref{eq:snr_net} and \ref{eq:snr_scale} and assuming coherent searches, the overall SNR for a network of $\mathcal{I}$ detectors with equal sensitivities is given by $\text{SNR}_\text{net} \propto \sqrt{\mathcal{I}}/D_L$. Assuming that GW sources are uniformly distributed across the sky, an $\mathcal{I}$-detector network can detect $\sqrt{\mathcal{I}}$ times further and up to $\sqrt{\mathcal{I}^3}$ more sources compared to a single detector network since the search volume scales as $V \propto D_L^3$.

\subsection{Sky coverage}
The sensitivity of a detector towards a particular sky location is determined by the antenna pattern in that given direction. Adding more detectors to the network at different geographical locations and orientations increases the sensitivity of the network to a wider region of the sky (increased sky coverage), consequently increasing the detection rate and volume along those directions \cite{2011CQGra..28l5023S}. 

Reference \cite{2012arXiv1202.4031M} presented a visual comparison between the network antenna pattern across the whole sky between a three-detector (HLV) network and a four-detector (HLKV) network, where `K' denotes KAGRA. As expected, results show that both networks are more sensitive to some regions in the sky than others. However, the HLKV network has higher overall network antenna power pattern and an overall increase in sky coverage is also reflected in the expansion of regions with relatively higher sensitivity.

\subsection{Observing time}
Adding detectors to the existing network also increases the duty cycle where two or more detectors are functional and simultaneously observing. This consequently increases the chances of the detectors picking up a coherent astrophysical signal and leading to higher detection rates \cite{2011CQGra..28l5023S}. 

\subsection{Sky localisation}

Sky localisation of a GW source is of vital importance for locating and identifying any existing electromagnetic counterparts to the GW event \cite{2018ApJ...854L..25P}. 
Ground-based GW detectors are nearly omnidirectional, so with a single detector we are not able to impose a strict constrains to the sky location of a GW event.
Nevertheless, sky localisation of GW signals improves significantly with multiple interferometers. The times of arrival at two detectors constrain the position of the source to an error ellipse in the sky map. Thus, having more detectors will reduce localisation volume by imposing stricter constraints to the location of the sources, improving the accuracy of locating the source in the sky \cite{2018MNRAS.479..601D, 2018ApJ...854L..25P}.
.

To sum up the points above, the advantages of having more detectors in the network include: (i) improvement in SNR and increased search volume, (ii) alignment-dependent sky coverage, (iii) increased rates of detection, and (iv) improved sky localisation.
\section{\textit{BayesWave} Overview} \label{sec:bw}

\textit{BayesWave} is a Bayesian data analysis algorithm that detects transient features in a stretch of detector data and identifies whether they are an astrophysical signal or instrumental noise. \textit{BayesWave} reconstructs non-Gaussian features in the data using a sum of sine-Gaussian (also called Morlet-Gabor) wavelets. The number of wavelets and their respective parameters are sampled using a trans-dimensional Markov Chain Monte Carlo algorithm, otherwise known as the Reversible-Jump Markov Chain Monte Carlo (RJMCMC). The RJMCMC is implemented to allow for adjustable number of wavelets and hence variable model dimensions. \textit{BayesWave} outputs posterior distributions and Bayesian evidences for three separate models: (i) Gaussian noise only, (ii) Gaussian noise with glitches and (iii) Gaussian noise with GW signal. The model evidences are then used for Bayesian model selection between the three scenarios.

\subsection{Wavelet Frames}
\label{subsec:Wavelets}

\textit{BayesWave} uses a sum of sine-Gaussian (also called Morlet-Gabor) wavelets to reconstruct non-Gaussian features (either signals or glitches) in the detector data. Even though Sine-Gaussian wavelets form a non-orthogonal frames\footnote{Discrete wavelets can form orthogonal bases for signal or glitch representations, but projecting the signal wavelets onto each detector requires the time translation operator which is computationally expensive. Despite the lack of orthogonality, sine-Gaussian wavelets are flexible in shape and have an analytic Fourier representation. Hence the analysis can be done in the frequency domain without the need of a time-translation operator. Further details can be found in Section 3 of \cite{2015CQGra..32m5012C}.}, their shape is variable in time-frequency plane and can optimally reconstruct a transient GW signal with no \emph{a priori} assumption on the signal source or morphology. 

The number of wavelets used in the reconstruction is marginalised via the RJMCMC, where signals with complex structure in time-frequency plane will use more wavelets in the reconstruction.
Previous studies \cite{2016PhRvD..94d4050L, 2018PhRvD..97j4057M} have shown that the number of wavelets scales linearly with SNR such that 
\begin{equation}
\label{eq:wavelet_complexity}
    N \approx \gamma + \beta \text{ SNR}
\end{equation}
where $\gamma$ and $\beta$ are constants which depend on waveform morphology. The results from Ref.~\cite{2016PhRvD..94d4050L} show that $\beta$ and hence $N$ increase with waveform complexity. For binary black hole (BBH) waveforms, the typical numbers are $\gamma=5.6$ and $\beta = 0.066$ for sine-Gaussian wavelet reconstructions \cite{2018PhRvD..97j4057M}.

In \textit{BayesWave}, each wavelet in the time domain has five intrinsic parameters $t_0, f_0, Q, A, \phi_0$ which represent central time, central frequency, quality factor, amplitude and phase offset respectively. These intrinsic parameters can be expressed as a single parameter vector $\vb*{\lambda}=\{t_0, f_0, Q, A, \phi_0 \}$ and the mathematical representation of a sine-Gaussian wavelet is given by
\begin{equation}
    \label{eq:wavelet}
    \Psi(t;t_0, f_0, Q, A, \phi_0) = Ae^{-\Delta t^2 / \tau ^2}\cos(2\pi f_0 \Delta t+\phi_0)
\end{equation}
with $\tau = Q/(2\pi f_0)$ and $\Delta t=t-t_0$ \cite{2016PhRvD..94d4050L}. 

The glitch model in \textit{BayesWave} is independent between detectors owing to the fact that noise artefacts are uncorrelated across different detectors. Hence, the set of glitch model parameters must contain the respective parameters for each individual detector across the network. The complete set of glitch model parameters for a network of detectors comprising Hanford, Livingston, and Virgo (HLV) can be written as \cite{2016PhRvD..94d4050L}
\begin{equation}
    \label{eq:glitch_param}
    \vb*{\theta_\mathcal{G}} = \{\vb*{\lambda}^H \cup \vb*{\lambda}^L \cup \vb*{\lambda}^V\}
\end{equation}
with $\vb*{\lambda}^i = \{\vb*{\lambda}_0^i \cup \vb*{\lambda}_1^i \cup \cdots \cup \vb*{\lambda}_{N^\mathcal{G}_i}^i\}$ where the numerical subscripts indicate a single wavelet used in the glitch model and $N^\mathcal{G}$ is the total number of wavelets in the glitch model. The superscripts indicates the $i$-th detector in the network.

In contrast to the glitch model, the signal model is common across all detectors in the network. As a result, signal models should have a single set of intrinsic wavelet parameters $\vb*{\lambda}^{\oplus}=\{\vb*{\lambda}_0 \cup \vb*{\lambda}_1 \cup \cdots \cup \vb*{\lambda}_{N^\mathcal{S}}\}$, along with a set of extrinsic parameter $\vb*{\Omega}= \{\alpha, \delta, \psi, \epsilon\}$ which sequentially describes the right ascension (RA), declination (dec), polarisation angle and ellipticity of the GW signal. The sky location (RA, dec) and polarisation angle of a source determine antenna beam patterns of the detector network, as well as provide information on the amplitude and the arrival-time delay of the signal in each detector \cite{2004ARNPS..54..525C}. Ellipticity defines the relative phase and amplitude of the plus and cross polarisations, $h_+$ and $h_\times$ respectively with $h_\times = \epsilon h_+ e^{i\pi /2}$. The ellipticity parameter, $\epsilon$ takes values between $0$ to $1$ with the lower and upper bounds denoting linear to circular polarisations respectively \cite{2015CQGra..32m5012C}. 
Altogether a complete set of signal model parameters is given by \cite{2016PhRvD..94d4050L}
\begin{equation}
    \label{eq:signal_param}
    \vb*{\theta_\mathcal{S}} = \{\vb*{\lambda}^\oplus \cup \vb*{\Omega}\}.
\end{equation}

\textit{BayesWave} produces posterior distributions of the parameters described above.  Each draw from the posterior contains a unique set of wavelet parameters (and extrinsic parameters for the signal model), which are then summed to produce a posterior on the waveform, $h(t)$.  By using this basis of sine-Gaussian wavelets, $h(t)$ is reconstructed with no \textit{a priori} assumption on the source of the GW signal.




\subsection{Model Selection}

In addition to waveform reconstruction, \textit{BayesWave} performs model selection between the signal and glitch hypotheses described above. The ratio of model evidences, otherwise known as the Bayes factor, is the key to model selection in Bayesian inference as it assesses the plausibility of two different models, $\mathcal{M}_\alpha$ and $\mathcal{M}_\beta$, parameterised by their respective parameter sets $\vec{\theta}_\alpha$ and $\vec{\theta}_\beta$. In other words, it quantifies which model is better supported by the data. The model evidence (also called the margainalised evidence) is given by
\begin{equation}
    \label{eq:model_evidence}
    p(\vec{s}|\mathcal{M}_\alpha) = \int p(\vec{\theta}_\alpha|\mathcal{M}_\alpha)p(\vec{s}|\vec{\theta}_\alpha, \mathcal{M}_\alpha) d\vec{\theta}_\alpha
\end{equation}
where $\vec{s}$ is the observed data, $\mathcal{M}_\alpha$ is the model, and $\vec{\theta}_\alpha$ is there parameter vector for model $\mathcal{M}_\alpha$. The prior probability of parameters $\vec{\theta}_\alpha$ before the data are observed is given by $p(\vec{\theta}_\alpha|\mathcal{M}_\alpha)$, and $p(\vec{s}|\vec{\theta}_\alpha,\mathcal{M}_\alpha)$ is the likelihood of obtaining the observed data $\vec{s}$, given the model $\mathcal{M}_\alpha$. Hence, the Bayes factor between models $\mathcal{M}_\alpha$ and $\mathcal{M}_\beta$, parameterised by their respective parameter vectors $\vec{\theta}_\alpha$ and $\vec{\theta}_\beta$ is
\begin{equation}
    \label{eq:bayesF}
    \begin{split}
    \mathcal{B}_{\alpha, \beta}(\vec{s}) &= \frac{p(\vec{s}|\mathcal{M}_\alpha)}{p(\vec{s}|\mathcal{M}_\beta)}.
    \end{split}
 \end{equation} 
$\mathcal{B}_{\alpha, \beta}(\vec{s}) > 1$ implies that model $\mathcal{M}_\alpha$ is more strongly supported by the data than model $\mathcal{M}_\beta$. To reduce computational costs, the \textit{BayesWave} algorithm calculates model evidence using thermodynamic integration \cite{2009PhRvD..80f3007L}. 

\textit{BayesWave} calculates the Bayes factor between the signal model (i.e. the data contains a real astrophysical signal), and the glitch model (i.e. the data contains an instrumetnal glitch). In Section~\ref{sec:BF_scaling} we discuss how the signal-to-glitch Bayes factor scales with SNR, the number of wavelets used in the MCMC, and the number of detectors in the network.

\subsection{Overlap}
In addition to distinguishing between signals and glitches, \textit{BayesWave} also produces a posterior distribution of the wavelet-expanded waveforms, $h(t)$ to match the true waveform, $h_s(t)$. One way to quantify the agreement or similarity between $h(t)$ and $h_s(t)$ is through the overlap, $\mathcal{O}$. 
Reconstructed waveforms in \textit{BayesWave} are analogous to waveform templates, hence the overlap between reconstructed models and the injected waveform can be computed the same way as the overlap in matched-filtering.

The normalised overlap between the two waveforms can be written as \cite{2017ApJ...839...15B}
\begin{equation}
\mathcal{O} = \frac{\inprod{h}{h_s}}{\sqrt{\inprod{h}{h}\inprod{h_s}{h_s}}}
\label{eq:overlap}
\end{equation}
where $(.|.)$ is the noise-weighted inner product as defined in Equation \ref{eq:noise_weighted_inprod}. Since Equation \ref{eq:overlap} is normalised, $\mathcal{O}$ takes values between $-1$ to $1$. When $\mathcal{O}=1$, there is a perfect match between the injected and recovered waveform; $\mathcal{O}=0$ implies that there is no match at all and $\mathcal{O}=-1$ implies a perfect anti-correlation.

Equation \ref{eq:overlap} only applies to a single detector. A network overlap, $\mathcal{O_{\text{net}}}$ is required to fully evaluate \textit{BayesWave}'s performance in recovering waveforms from all the detectors combined. In order to define the network overlap, we sum each factor in Equation \ref{eq:overlap} over all $\mathcal{I}$ detectors in the network such that
\begin{equation}
    \mathcal{O_{\text{net}}} = \frac{\sum_{i=1}^{\mathcal{I}} \inprod{h^{(i)}}{h_s^{(i)}}}{\sqrt{\sum_{i=1}^{\mathcal{I}}\inprod{h^{(i)}}{h^{(i)}}\sum_{i=1}^{\mathcal{I}}\inprod{h_s^{(i)}}{h_s^{(i)}}}},
    \label{eq:network_overlap}
\end{equation}
where $h^{(i)}$ and $h_s^{(i)}$ denote the recovered waveform and waveform present in detector $i$ respectively. 

\section{Analytic Bayes Factor Scaling}
\label{sec:BF_scaling}

In this work, we aim to understand the behavior of the Bayes factor between signal and glitch models for networks comprising different numbers of GW detectors. Hence it is in our interest to analytically understand the conditions of model selection. We want to know under what circumstances a model is favoured over another.

\subsection{Occam Penalty}
\label{subsec:occam_penalty}
A key to understanding Bayes factor behavior when using a trans-dimensional model as \textit{BayesWave} does, is the role of the Occam penalty.

The parameter value at which the posterior distribution peaks is known as the maximum a posteriori (MAP) value, denoted as $\vec{\theta}_\text{MAP}$. For high SNR events, the integrand of model evidence in Equation \ref{eq:model_evidence} peaks sharply in the vicinity of the MAP. Following the Laplace-Fisher approximation, the integral can be estimated as
\begin{equation}
    \label{eq:laplace_appx}
    p(\vec{s}|\mathcal{M}) \simeq p(\vec{s}|\vec{\theta}_{\text{MAP}}, \mathcal{M})p(\vec{\theta}_{\text{MAP}}| \mathcal{M})(2\pi)^{D/2}\sqrt{\text{det}C}.
\end{equation}
where $p(\vec{s}|\vec{\theta}_{\text{MAP}}, \mathcal{M})$ is the MAP likelihood; $p(\vec{\theta}_{\text{MAP}}| \mathcal{M})$ is the prior evaluated at the MAP parameter values; $D$ is the dimension of the model; and $\det C$ is the determinant of the full covariance matrix for the $N$ wavelets used in waveform reconstruction. If the covariance matrix for a single wavelet is $C_n$, then we have
\begin{equation}
\det C = \prod^N_{n=1} \det C_n,
\end{equation}
assuming minimal overlap between the wavelet parameter spaces. Since the Laplace-Fisher approximation is associated with the MAP likelihood, the covariance matrix can be approximated as the inverse of the Fisher Information Matrix (FIM), $\Gamma$ \cite{doi:10.2307/3315678}. A comprehensive discussion of the FIM and its relation to wavelet parameter jump proposal is presented in Appendix \ref{app:FIM}. 

By definition, $\det C$ measures the variance of the likelihood. Thus, $\sqrt{\det C}$ quantifies the characteristic spread of the likelihood function. The product of $\sqrt{\det C}$ and $(2\pi)^{D/2}$, which account for the dimensionality of the model, can then be used as a measure for the volume of the uncertainty ellipsoid (posterior volume), $\Delta V_\mathcal{M}$ for a given model $\mathcal{M}$ \cite{2016PhRvD..94d4050L, 2017LRR....20....2R, littenberg...2009}. Assuming uniform priors for all wavelet parameters, one can also write $p(\vec{\theta}_{\text{MAP}}| \mathcal{M}) = 1/V_\mathcal{M}$ where $V_\mathcal{M}$ represents the total parameter space volume. Hence, the last three factors of Equation \ref{eq:laplace_appx} can collectively be interpreted as the fraction of the prior occupied by the posterior distribution, such that the model evidence is now given by
\begin{equation}
    \label{eq:occam_fac}
    p(\vec{s}|\mathcal{M}) \simeq p(\vec{s}|\vec{\theta}_{\text{MAP}}, \mathcal{M})\frac{\Delta V_{\mathcal{M}}}{{V_{\mathcal{M}}}}.
\end{equation} 
where $\Delta V_{\mathcal{M}}/V_\mathcal{M}$ is the ``Occam penalty factor''.  

Following equations \ref{eq:bayesF}  and \ref{eq:occam_fac}, the Bayes factor between two models can be re-expressed as
\begin{equation}
\mathcal{B}_{\alpha, \beta}(\vec{s}) = \Lambda_{\alpha, \beta}(\vec{s})\frac{\Delta V_{\alpha}}{V_\alpha}\frac{V_\beta}{\Delta V_\beta}
\label{eq:BF_2}
\end{equation}
where the ratio of MAP likelihoods is given by
\begin{align}
    \Lambda_{\alpha, \beta}(\vec{s})
    &= \frac{p(\vec{s}|\vec{\theta}_{\text{MAP},\alpha})}{p(\vec{s}|\vec{\theta}_{\text{MAP},\beta})}.
    \label{eq:map-likelihood}
\end{align}
Equation \ref{eq:BF_2} suggests that the Bayes factor is dependent on the likelihood ratio and the ratio of the Occam penalty factors. The Occam factor penalises models that require an unnecessarily large parameter space volume to fit the data by suppressing the model evidence. 
Note that Occam penalty is not an intentionally added component to the Bayes factor, rather it is inherently imposed as a result of using the Bayes Theorem.

As a heuristic explanation as to how the Occam penalty aids in \textit{BayesWave}'s ability to distinguish between signals and glitches, recall that signal models ($\mathcal{S}$) for each detector share the same intrinsic parameters and four extrinsic parameters. Since there are five intrinsic parameters ($t_0, f_0, Q, A, \phi_0$) per wavelet, the dimension of signal models scales as
\begin{equation} 
D_\mathcal{S} \sim 5N + 4
\end{equation}
where $N$ is the number of wavelets. Glitch models ($\mathcal{G}$), on the other hand, have no extrinsic parameters but the glitch model of each detector is described by a unique set of intrinsic parameters. Assuming that signal and glitch models use the same number of wavelets such that $N^\mathcal{G} =  N^\mathcal{S} = N$ (see Appendix \ref{app:BF_scaling}), the dimension of glitch models scales as \cite{2015CQGra..32m5012C}
\begin{equation}
    D_\mathcal{G} \sim 5N\mathcal{I}.
\end{equation}
One therefore has $D_\mathcal{G} > D_\mathcal{S}$ for $\mathcal{I}\geq 2$. This implies that the total parameter space volume for the glitch model is larger than that of the signal model (i.e. $V_\mathcal{G} > V_\mathcal{S}$). If both models fit the data equally well (i.e. $\Lambda_{\mathcal{S}, \mathcal{G}}\approx1$ and $\Delta V_\mathcal{S}\approx V_\mathcal{G}$), then by Occam's razor we should expect to see a selection bias towards the signal model as $\mathcal{I}$ increases. In other words, Equation \ref{eq:BF_2} gives
\begin{equation}
\mathcal{B}_{\mathcal{S}, \mathcal{G}}(\vec{s}) = \Lambda_{B_{\mathcal{S}, \mathcal{G}}}(\vec{s})\frac{\Delta V_{\mathcal{S}}}{\Delta V_\mathcal{G}}\frac{V_\mathcal{G}}{V_\mathcal{S}} > 1
\end{equation}
with increasing $\mathcal{I}$.

In Section \ref{subsec:BF_N}, we use the Laplace approximation to the Bayesian evidence to derive an analytic scaling of the Bayes factor.

\subsection{Dependence of Bayes factor on number of detectors}
\label{subsec:BF_N}
In Ref.~\cite{2016PhRvD..94d4050L}, Littenberg et al. put forth an analytic scaling of the log signal-to-glitch Bayes factor, $\ln\mathcal{B}_{\mathcal{S},\mathcal{G}}$, in an effort to fully understand \textit{BayesWave}'s ability to robustly distinguish astrophysical signals from instrumental glitches.  They showed that the primary scaling of the Bayes factor goes as \begin{equation}
\ln\mathcal{B}_{\mathcal{S},\mathcal{G}}\propto N \ln(\text{SNR}_\text{net})
\label{eq:BSG_previous}
\end{equation}
where $N$ is the number of wavelets used in the reconstruction, which is related to the signal morphology and SNR as described in Equation~\ref{eq:wavelet_complexity}. 
The dependence of Bayes factor on $N$ (and therefore the complexity of the signal in time frequency plane) differentiates \textit{BayesWave} from other unmodelled searches whose detection statistics scale primarily with SNR. The scaling found in Ref.~\cite{2016PhRvD..94d4050L} assumes a network comprising two GW detectors. Here we extend this work to an arbitrary number of detectors $\mathcal{I}$.

We begin with the Laplace approximation of model evidences for the signal and glitch models. From equation \ref{eq:laplace_appx}, we find
\begin{widetext}
\begin{equation}
\ln p\inprod{d}{\mathcal{S}} \simeq  \frac{\text{SNR}_\text{net}^2}{2} - \frac{5N^\mathcal{S}}{2} - N^\mathcal{S}\ln(V_\lambda) + \sum _{n=1} ^{N^\mathcal{S}} \ln \left( \frac{\bar{Q}_n}{\text{SNR}_{\text{net},n}^5}\right) + \frac{D_\Omega}{2} + \ln \frac{\sqrt{\det C_\Omega}}{V_\Omega}
\label{eq:evidence_s}
\end{equation}

\begin{equation}
\ln p\inprod{d}{\mathcal{G}} \simeq  \frac{\text{SNR}_\text{net}^2}{2} - \sum_{i=1}^\mathcal{I} \left[\frac{5N_i^\mathcal{G}}{2} + N_i^\mathcal{G}\ln(V_\lambda) - \sum _{n=1} ^{N_i^\mathcal{G}} \ln 
\left( \frac{\bar{Q}_n}{\text{SNR}_{i,n}^5}\right)\right]
\label{eq:evidence_g}
\end{equation}
\end{widetext}
with $\bar{Q}_n \equiv (2\pi)^{\sfrac{5}{2}}\frac{\sqrt{2}Q_n}{\pi}$. $V_\lambda$ is the prior volume of intrinsic parameters and $N^x_i$ is the total number of wavelets for model $x$. The subscript $i$ refers to detector $i$ in the network and $n$ labels an individual wavelet from the set of wavelets for a given model. For instance: $\text{SNR}_{i, n}$ is the SNR of wavelet $n$ in the $i$-th detector\footnote{Each individual wavelet used in signal or glitch model reconstruction has an amplitude which can be converted into to SNR. For details, see \cite{2015CQGra..32m5012C}.}. In the last two terms of Equation \ref{eq:evidence_s}, $D_\Omega$=4, $C_\Omega$ and $V_\Omega$ denote the dimension, covariance matrix and the prior volume of extrinsic parameters respectively. The full derivation from Equation \ref{eq:laplace_appx} to Equations \ref{eq:evidence_s} and \ref{eq:evidence_g} can be found in Section III(A) of Ref. \cite{2016PhRvD..94d4050L}.

To simplify the expressions for these evidences, we follow the same assumptions used in Ref~\cite{2016PhRvD..94d4050L}, and which are detailed further in Appendix~\ref{app:BF_scaling}.  One simplifying assumption we highlight here again is that the number of wavelets used in the signal model will be approximately the same as the glitch model, and so we set $N^\mathcal{S}=N^\mathcal{G}\equiv N$ (i.e. the $N$ in Equation~\ref{eq:BSG_previous}).
Upon implementing the assumptions in Appendix \ref{app:BF_scaling}, the theoretical log Bayes factor between the signal and glitch model for a network of $\mathcal{I}$ detector(s) is given by $\ln \mathcal{B}_{\mathcal{S},\mathcal{G}} \simeq \ln p\inprod{d}{\mathcal{S}} - \ln p\inprod{d}{\mathcal{G}}$:
\begin{widetext}
\begin{align}
    \ln \mathcal{B}_{\mathcal{S},\mathcal{G}} \simeq
    (\mathcal{I}-1)\left[\frac{5N}{2} + N\ln(V_\lambda) - \sum _{n=1} ^{N} \ln \left( \bar{Q}_n\right) + 5N\ln \left(\frac{\text{SNR}_\text{net}}{\sqrt{N}}\right)\right] - \frac{5}{2}\mathcal{I}N\ln(\mathcal{I}) +\left(2 + \ln \frac{\sqrt{\det C_\Omega}}{V_\Omega}\right).
    \label{eq:BF_scaling}
\end{align}
\end{widetext}
The equation shows explicit dependence of the Bayes factor on network SNR, number of wavelets and number of detectors. We pay close attention to the scaling 
\begin{equation}
\ln \mathcal{B}_{\mathcal{S},\mathcal{G}} \propto \mathcal{I}N\ln\text{SNR}_{\text{net}}
\label{eq:BF_scaling2}
\end{equation}
which now has an extra scaling factor of $\mathcal{I}$ compared to Equation \ref{eq:BSG_previous}. 

The dependence on the number of wavelets used implies that the signal model is favoured over the glitch model with increasing waveform complexity (higher $N$). In other words, a more complex waveform is more likely to be classified as a signal \cite{2016PhRvD..94d4050L}. This analytic result agrees with the discussion in Section \ref{subsec:occam_penalty} where if two models fit the data equally well, the less complex model will be selected to represent the waveform. The proportionality $\ln \mathcal{B}_{\mathcal{S},\mathcal{G}} \propto \mathcal{I}$ suggests that for signals with equal SNR and $N$, the Bayes factor should increase if we increase the number of detectors in the network.  Again, this result agrees with the discussion in Section~\ref{subsec:occam_penalty}; including more detectors in the network increases the dimensionality of the glitch model and thus the signal model will be even more strongly preferred.


\section{Injection data set} \label{sec:method}

\begin{figure}[t]
\centering
\includegraphics[width=0.49\textwidth]{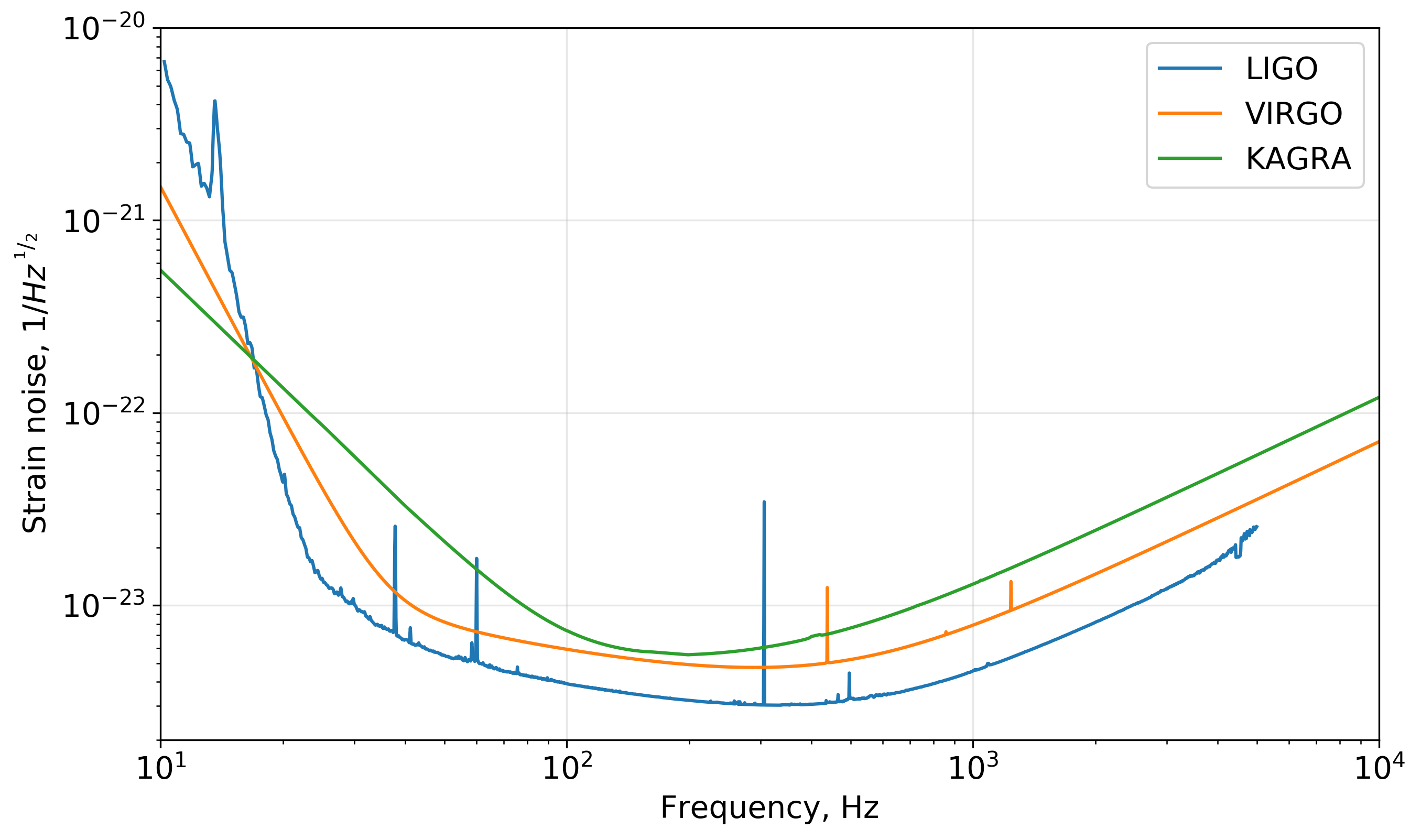}
    \caption{Projected LIGO, Virgo and KAGRA strain noise (i.e. amplitude spectral density), $\sqrt{S_n}$ as a function of frequency for the fourth observing run, O4. The data used to generate the noise curves above are retrieved from \cite{2018LRR....21....3A}.}
    \label{fig:PSD}
\end{figure}

To empirically test the Bayes factor scaling given by Equation \ref{eq:BF_scaling}, as well as investigate the effect on waveform reconstructions with detector networks of different sizes, we inject a set of simulated BBH signals into simulated detector noise and recover them using \textit{BayesWave}. While \textit{BayesWave} is a flexible algorithm that can detect a variety of signals from different sources, we use BBH waveforms as our test bed because they are well-understood sources, and have previously been used to study the performance of BayesWave \cite{2016PhRvD..94d4050L,2017ApJ...839...15B, 2020arXiv200309456G}.

In this work, we use tools from the LIGO Analysis Library \citep{lalsuite} to inject a set of non-spinning binary black holes (BBH) with equal component masses of $30M_\odot$.  We use the phenomenological waveform \texttt{IMRPhenomD} to model spinning but non-precessing binaries using a combination of analytic post-Newtonian (PN), effective-one-body (EOB) and numerical relativity (NR) methods \cite{2016PhRvD..93d4006H, 2016PhRvD..93d4007K}.  The GW sources are distributed isotropically across the sky, and the inclinations $\iota$ are distributed uniformly in $\arccos{\iota}$. ${\rm SNR}_{\rm net}$ is distributed uniformly in SNR$_\text{net}\in \{10,50\}$ where this SNR is calculated from a network comprising the HL detectors.  

We inject 150 BBH signals into Gaussian noise coloured by the projected PSD of LIGO, Virgo and KAGRA for the fourth observing run, O4, as given in the LIGO, Virgo and KAGRA Observing Scenario \cite{2018LRR....21....3A}. The noise curves are shown in Figure \ref{fig:PSD}.

We then recover the injected signals with \textit{BayesWave} in three different scenarios: (i) Running only on Hanford and Livingston (HL) data (a two detector network), (ii) Running on the Hanford, Livingston, and Virgo (HLV) data (a three detector network) and (iii) Running on the Hanford, Livingston, KAGRA and Virgo (HLKV) data (a four detector network).  All three detector configurations use the exact same injection data set.

In the two following sections, Sections \ref{sec:BayesFactor} and \ref{sec:overlap}, we analyse two figures of merit: (i) Bayes factor and (ii) the overlap. By comparing these quantities between the HL and HLV networks, we can evaluate the performance of \textit{BayesWave} in recovering the injected waveforms from detector networks of different sizes. As an extension to previous studies on sky localisation with expanded detector networks, we also compare the accuracy of BayesWave in recovering the sky location from detector networks of different sizes in Section \ref{sec:skyloc}. 

\section{Results}
\subsection{Recovered Bayes factors} 
\label{sec:BayesFactor}

\begin{figure*}[t]
\centering
\begin{minipage}{0.49\textwidth}
\includegraphics[width=\textwidth]{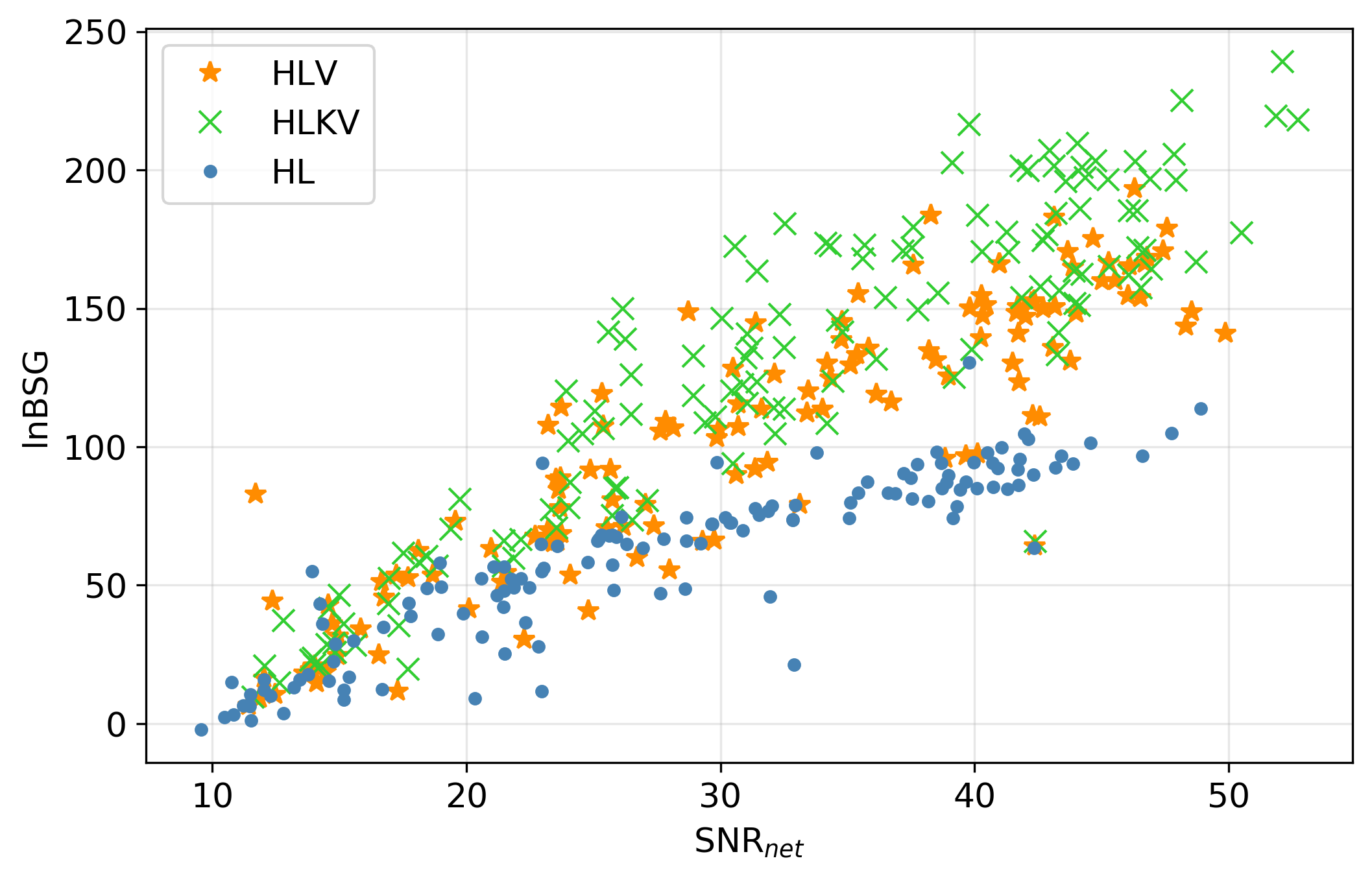}
\end{minipage}\hfill
\begin{minipage}{0.49\textwidth}
\includegraphics[width=\textwidth]{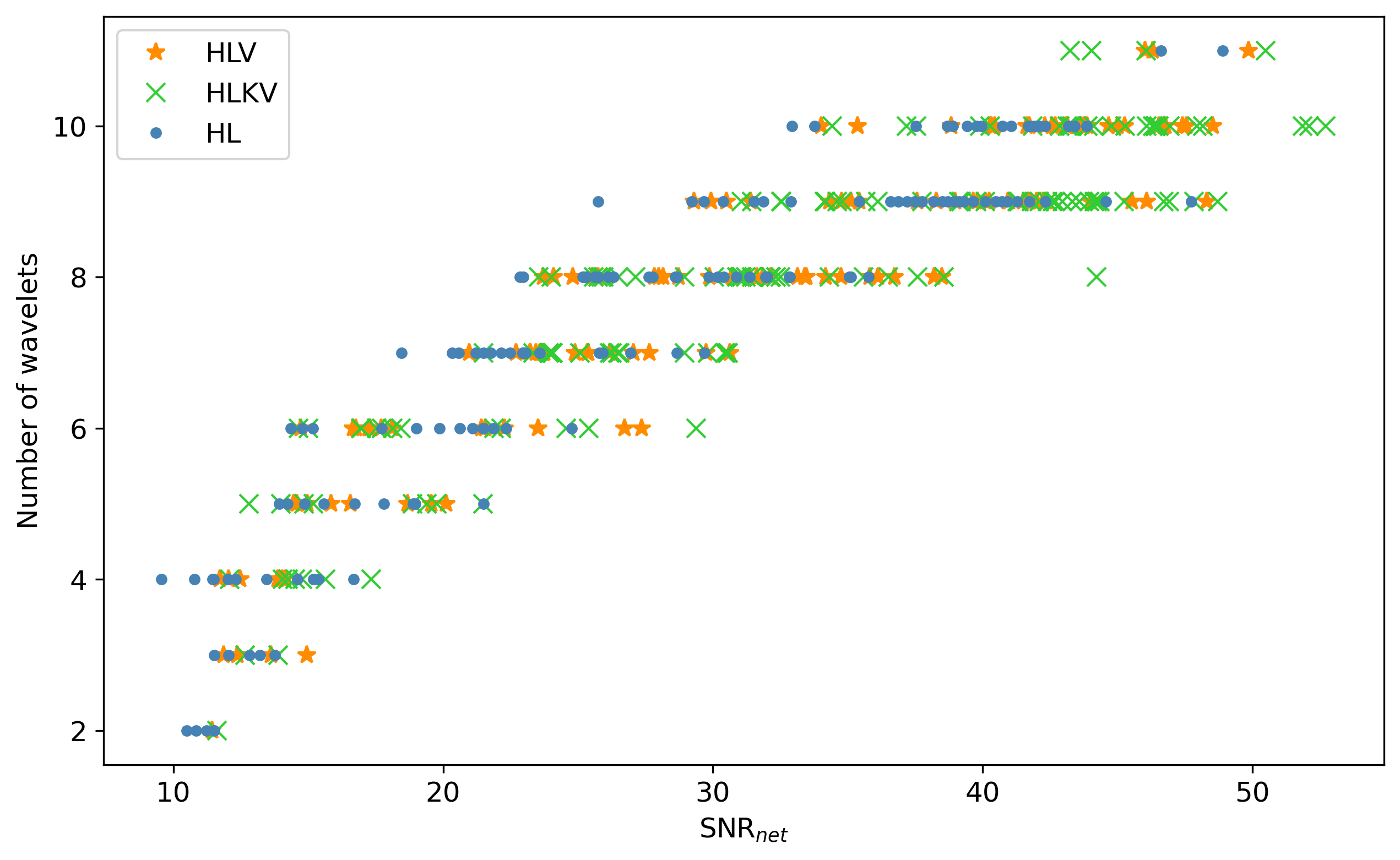}
\end{minipage}\par
\vskip\floatsep
\includegraphics[width=0.68\textwidth]{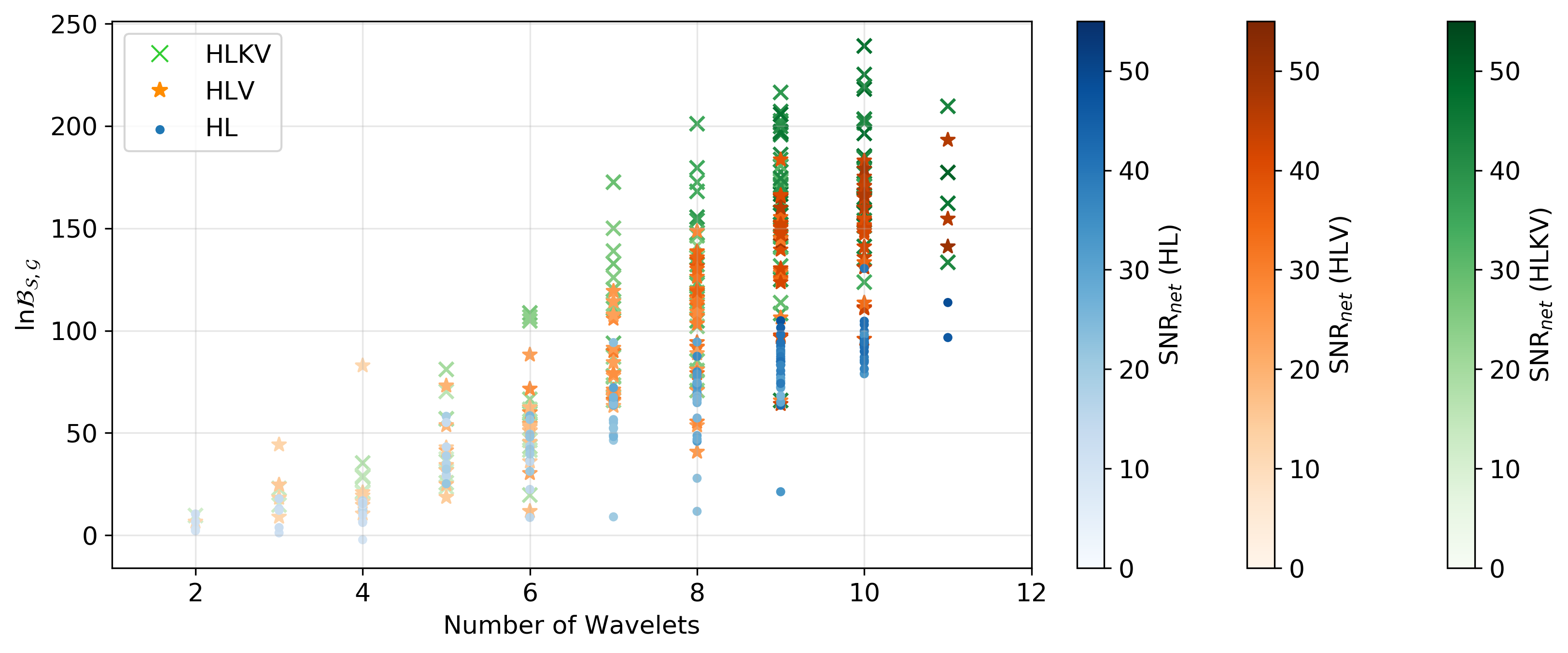}
    \caption{Top left panel shows the log signal-to-glitch Bayes Factor $\ln \mathcal{B}_{\mathcal{S},\mathcal{G}}$ of BBH injection recoveries versus network signal-to-noise ratio, SNR$_{\text{net}}$. Each data point represents one BBH injection. Top right panel shows the median number of wavelets used in signal model reconstruction for each injection, $N$ versus SNR$_{\text{net}}$. Bottom panel shows $\ln \mathcal{B}_{\mathcal{S},\mathcal{G}}$ versus $N$, and the three colour bars indicates the network SNR of each data point in the corresponding detector network. In the top panels, the horizontal axis corresponds three different network SNRs: (i) for the blue dot data points it corresponds to SNR$_{\text{net}}$ of the HL network, (ii) for the orange star data points it corresponds to SNR$_{\text{net}}$ of the HLV network, (iii) for the green cross data points it corresponds to SNR$_{\text{net}}$ of the HLKV network.}
    \label{fig:BSG}
\end{figure*}

After analysing the injections described in Section~\ref{sec:method}, we use the model evidences calculated by \textit{BayesWave} to understand the impact of GW detector network size on the log signal-to-glitch Bayes factor, $\ln \mathcal{B}_{\mathcal{S},\mathcal{G}}$.
For all the analyses in this paper, we only injections that have been identified as inconsistent with Gaussian noise (this can be either a signal or glitch) by \textit{BayesWave}. Injections indicated to be consistent with the Gaussian noise model ($\mathcal{N}$) by \textit{BayesWave} are removed from the data set, since it would be meaningless to evaluate their respective signal and glitch model evidences. In other words, injections with $\ln \mathcal{B}_{\mathcal{S},\mathcal{N}}$ error bars encompassing values below zero are removed from the data set. The widths of $\ln \mathcal{B}_{\mathcal{S},\mathcal{N}}$ error bars are given by \cite{2015CQGra..32m5012C}
\begin{widetext}
\begin{equation}
    \Delta [\ln \mathcal{B}_{\mathcal{S},\mathcal{N}}] = \sqrt{\{\Delta[\ln p\inprod{d}{\mathcal{S}}]\}^2 + \{\Delta[\ln p\inprod{d}{\mathcal{N}}]\}^2}
\end{equation}
\end{widetext}
where $\Delta[\ln p\inprod{d}{\mathcal{M}}]$ is the uncertainty for the logarithmic evidence of model $\mathcal{M}$. A total of 14 data points are removed under this constraint.  These events are all low SNR$_\text{net}$ injections.

The top left panel of Figure \ref{fig:BSG} shows $\ln \mathcal{B}_{\mathcal{S},\mathcal{G}}$ as a function of SNR$_{\text{net}}$ for the HL, HLV and HLKV networks. All three networks show a clear trend of increasing Bayes Factor with increasing network SNR as expected. Our results also show that the HLKV injections have the highest SNR overall, agreeing with Equation \ref{eq:snr_net} which indicates that increasing $\mathcal{I}$ increases $\mathrm{SNR}_\mathrm{net}$. Furthermore, we can see that injections at comparable SNRs are recovered with higher $\ln\mathcal{B}_{\mathcal{S},\mathcal{G}}$ in the HLV network than the HL network. In other words, even after accounting for the increased SNR, we observe further enhancement in detection confidence for an expanded detector network, suggesting that $\ln\mathcal{B}_{\mathcal{S},\mathcal{G}}$ is related to $\mathcal{I}$, and not just the SNR of the signal as predicted by Equation \ref{eq:BF_scaling}. 



 The top right panel of Figure~\ref{fig:BSG} shows the median number of wavelets used in the \textit{BayesWave} reconstruction, $N$ versus the injected SNR in the respective detector networks, SNR$_{\text{net}}$. The median here refers to the median of posterior distribution for $N$.
We see that $N$ increases systematically with SNR$_{\text{net}}$ in both the HL and HLV networks. This is expected since the detectors are able to pick up more complex features of the waveform at high SNR. At low SNR (SNR $\lesssim 15$) there is a slight deviation from the linear trend described by Equation~\ref{eq:wavelet_complexity} between $N$ and SNR in both detector networks. This is primarily due to the prior on the number of wavelets.  This prior is determined empirically from runs in LIGO data after O1, and peaks around $N=3$~\cite{BWIII}. 
$N$ also depends on waveform morphology and complexity \cite{2016PhRvD..94d4050L, 2018PhRvD..97j4057M}. Injecting the same set of BBH waveforms into all three detector configurations result in similar trends between $N$ and SNR$_\text{net}$.

Equation \ref{eq:BF_scaling} shows that $\ln\mathcal{B}_{\mathcal{S},\mathcal{G}}$ also scales with the number of wavelets used in the reconstruction. Hence we also show empirically how the dimensionality of signal model (i.e. the number of wavelets) also contributes to the increase in $\ln\mathcal{B}_{\mathcal{S},\mathcal{G}}$ for different $\mathcal{I}$. We show this in the bottom panel of Figure \ref{fig:BSG} by plotting $\ln \mathcal{B}_{\mathcal{S},\mathcal{G}}$ versus $N$. Colour bars indicate the SNR$_{\text{net}}$ of each data point. For all three detector configurations, $\ln \mathcal{B}_{\mathcal{S},\mathcal{G}}$ generally increases with $N$, as predicted by Equation~\ref{eq:BF_scaling}. At low SNRs (i.e. SNR$<15$), detector networks recover the waveform with $N\leq 3$ and $\ln \mathcal{B}_{\mathcal{S},\mathcal{G}} \leq 50$ because low SNR injections have low amplitude features which are harder to reconstruct resulting in lower detection confidence.
It is clear for injections recovered with  $N>3$ that 
$\ln \mathcal{B}_{\mathcal{S},\mathcal{G}}$ in the HLKV network are generally higher than that of the HL and HLV networks at comparable $N$ and SNR$_\text{net}$. This again emphasizes the point that the Bayes factor scales with $\mathcal{I}$. 

A more thorough investigation of the relation between the empirical and analytic Bayes factor can be found in Appendix \ref{app:SG_injections}, where we use a simplified injection set of single sine-Gaussian wavelets. By recovering sine-Gaussian wavelets with sine-Gaussian wavelets, Equation~\ref{eq:wavelet_complexity} reduces to $N=1$. The results show that the empirical scaling of the Bayes factor with $\mathcal{I}$ agrees with the analytical scaling in Equation \ref{eq:BF_scaling} to a good approximation.

In summary, we show by comparing Bayes Factors between the HL, HLV and HLKV networks that expanding detector networks increases detection confidence. Our empirical results are consistent with the analytic results discussed Section \ref{sec:BF_scaling}, viz. $\ln \mathcal{B}_{\mathcal{S},\mathcal{G}}\propto\mathcal{I}N\ln\mathrm{SNR}_\mathrm{net}$. Heuristically, this can be understood via Occam's razor: if coincident identical glitches are unlikely in two detectors, they are even more unlikely in three or more detectors. Therefore when identical waveforms are detected simultaneously across larger networks, they have a higher likelihood of being a signal.

\begin{figure}[t]
\centering
\includegraphics[width=.49\textwidth]{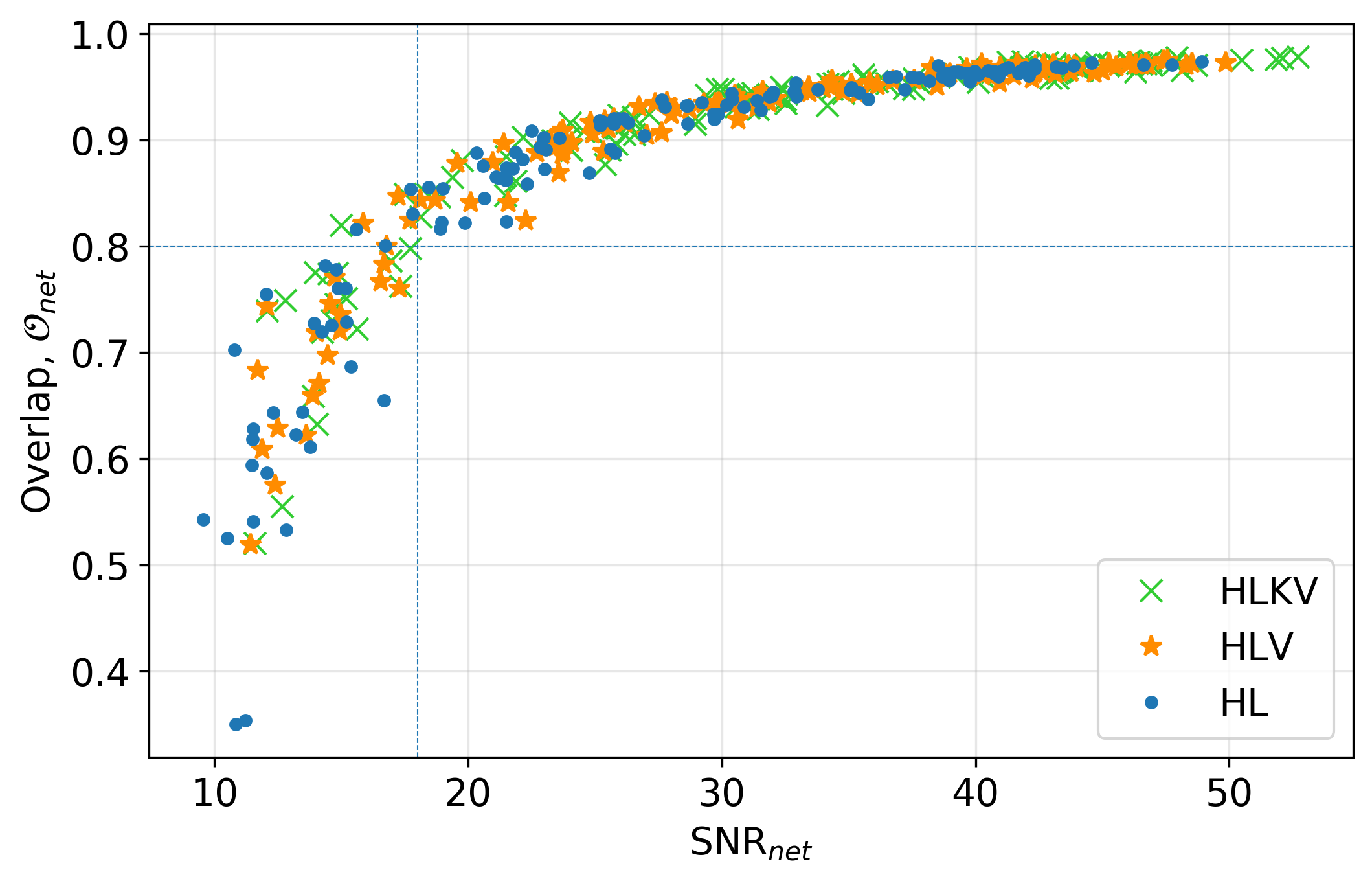}
    \caption{Median overlap between the injected and recovered waveform, $\mathcal{O}_\text{net}$ of the HL (blue dot) and HLV (orange star) network, as a function of SNR$_\text{net}$. The horizontal blue line indicates $\mathcal{O}_\text{net}=0.8$ and the vertical blue line indicates SNR$_\text{net}>15$.}
    \label{fig:overlap}
\end{figure}

\begin{figure*}[t]
\centering
\includegraphics[width=\textwidth]{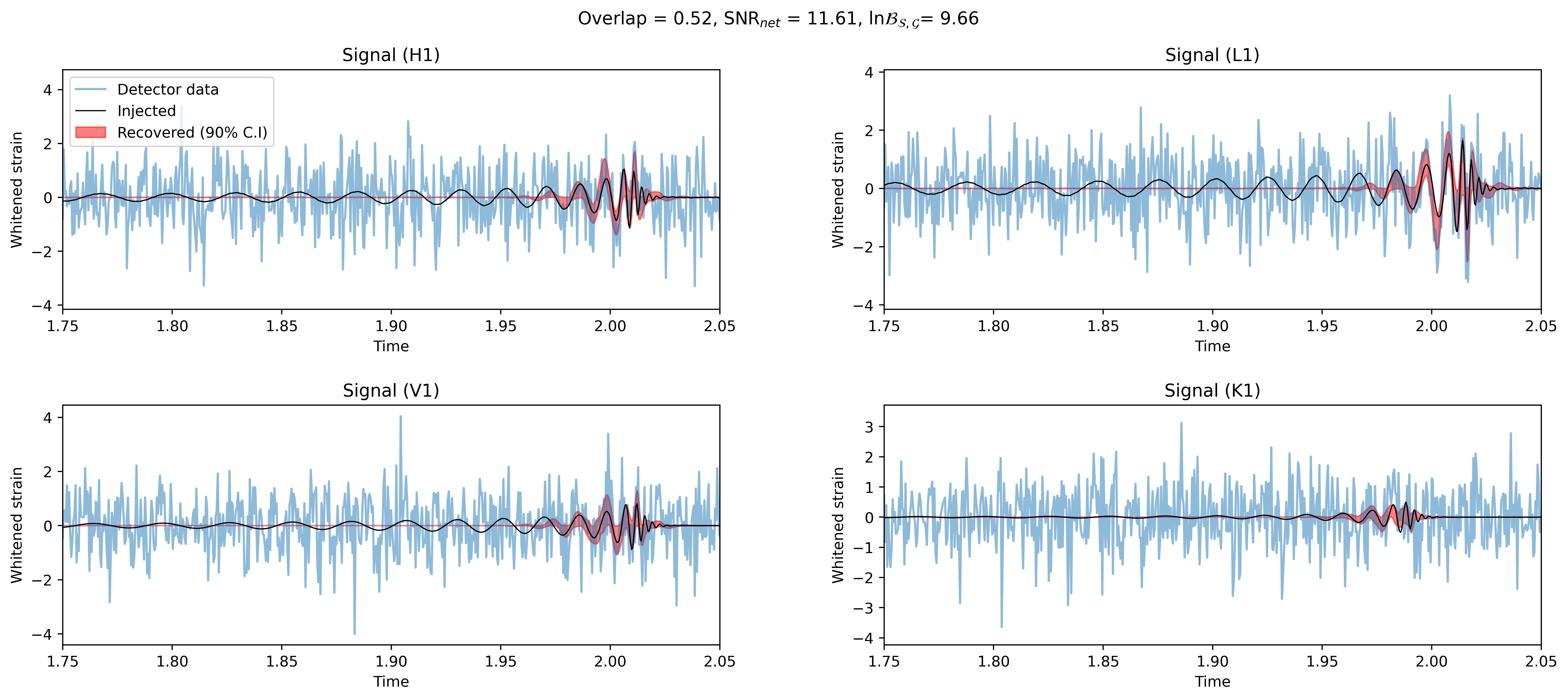}\par
\vskip\floatsep
\includegraphics[width=\textwidth]{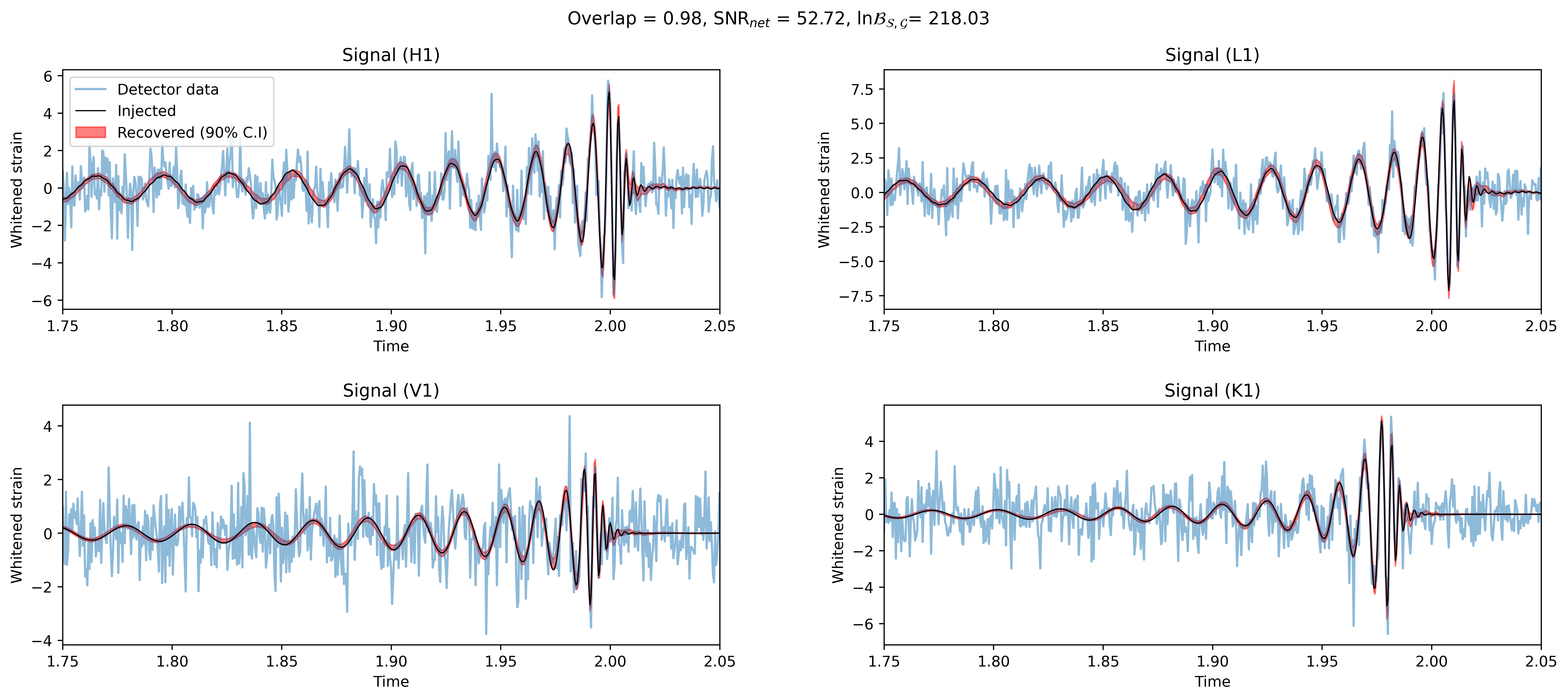}
    \caption{The top panel shows, for an injection with SNR$_{\text{net}}=11.61$ and $\mathcal{O}=0.52$, the injected waveform (black), the detector data (blue) and the $90\%$ credible interval of the recovered waveform (red) for each detector in the HLKV network. Similarly in the bottom panel but for an injection with SNR$_{\text{net}}=52.72$ and $\mathcal{O}=0.98$.}
    \label{fig:overlap_visual}
\end{figure*}

\subsection{Recovered Waveform Overlap} 
\label{sec:overlap}

In the previous section, we showed that for a set of BBH waveforms, $\ln \mathcal{B}_{\mathcal{S},\mathcal{G}}$ increases with a larger number of detectors in the network, meaning with more detectors our confidence in detection is strengthened. In this section, we quantify the accuracy of \textit{BayesWave} in waveform recovery by comparing the overlap (also sometimes called the match) between the injected and recovered waveforms for the HL, HLV and HLKV detector networks. The network overlap, $\mathbb{O}_{net}$ is given by Equation \ref{eq:network_overlap}. For the rest of this paper, any mention of overlap refers to the network overlap.

Figure \ref{fig:overlap} shows the median overlap, $\mathbb{O}_{net}$ as a function of network SNR, where $\mathbb{O}_{net}$ of all three detector networks show positive correlation with their respective network SNR. This observation is consistent with previous results, which show that network overlap scales with SNR \cite{2018PhRvD..97j4057M, 2020arXiv200309456G}. To illustrate how waveform reconstruction improves with SNR, Figure \ref{fig:overlap_visual} shows the injected waveform (black), the detector data (blue) and the $90\%$ credible interval of the recovered waveform (red) for two events in the HLKV network. The top and bottom panels show the waveforms for the injection recovered with the smallest overlap ($\mathbb{O}_\text{min} = 0.52$) and largest overlap ($\mathbb{O}_\text{max} = 0.98$) of the whole injection data set respectively.  The event with the smallest overlap has SNR$_\text{net}=11.6$ and was recovered with $\ln\mathcal{B}_{\mathcal{S},\mathcal{G}}=9.66$, while the event with the largest overlap has SNR$_\text{net}=52.72$ and was recovered with $\ln\mathcal{B}_{\mathcal{S},\mathcal{G}}=218.0$
This is consistent with the observed trend between overlap and network SNR in Figure \ref{fig:overlap}. The similar trend between overlap and network SNR between all three detector configurations indicates that waveform reconstruction fidelity is not directly related to the number of detectors in the network. 

However as noted earlier, increasing the number of detectors does increase the network SNR.  
By comparing the percentage of waveforms recovered with overlap above a given threshold for all three detector configurations, we show that having an additional detector allows us to better reconstruct the signal waveform. The threshold is arbitrarily defined here to be $\mathbb{O}_\text{net}>0.8$ and is indicated by the horizontal blue line in Figure \ref{fig:overlap}. We found that $81\%$ of the injections were recovered with $\mathbb{O}_\text{net}>0.8$  for the HL network, $86\%$ for the HLV network and $87\%$ for the HLKV network. 

While the inclusion of additional detector(s) does not have an extra benefit in the same way it does for the Bayes factor as shown in the previous section, it nonetheless allows us to better reconstruct the signal waveform due to increased SNR. However, the improvement is less significant upon the addition of KAGRA, since it is less sensitive compared to Virgo as shown in Figure \ref{fig:PSD} and therefore the increase in SNR is less compared to when Virgo is added to the network. The overall results also show that \textit{BayesWave} is able to reconstruct waveforms reasonably well with all three detector configurations for injections with SNR$_\text{net}>18$ as indicated by the vertical blue line in Figure \ref{fig:overlap}. 



\begin{figure*}[t]
\centering
\includegraphics[width=0.95\textwidth]{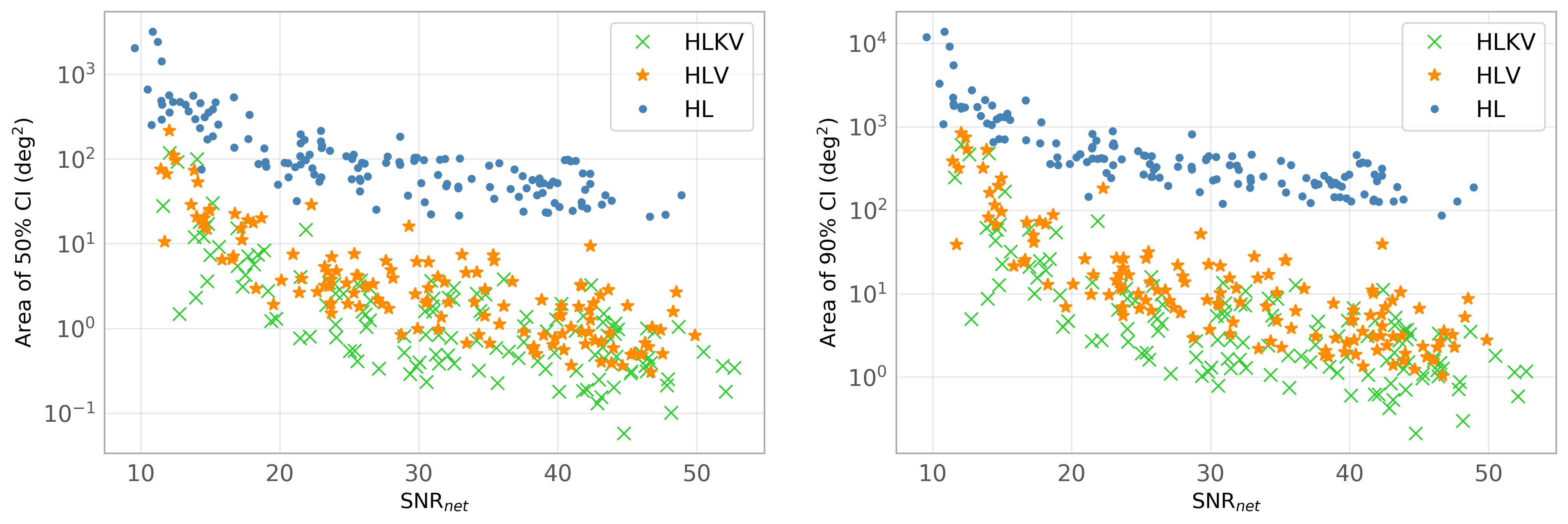}
    \caption{The left panel shows the sky area enclosed within the $50\%$ credible interval (CI) in square degrees versus the network SNR of the corresponding detector network. Similarly on the right panel, except for the $90\%$ CI.}
    \label{fig:skyloc}
\end{figure*}

\begin{figure}[t]
\centering
\includegraphics[width=.48\textwidth]{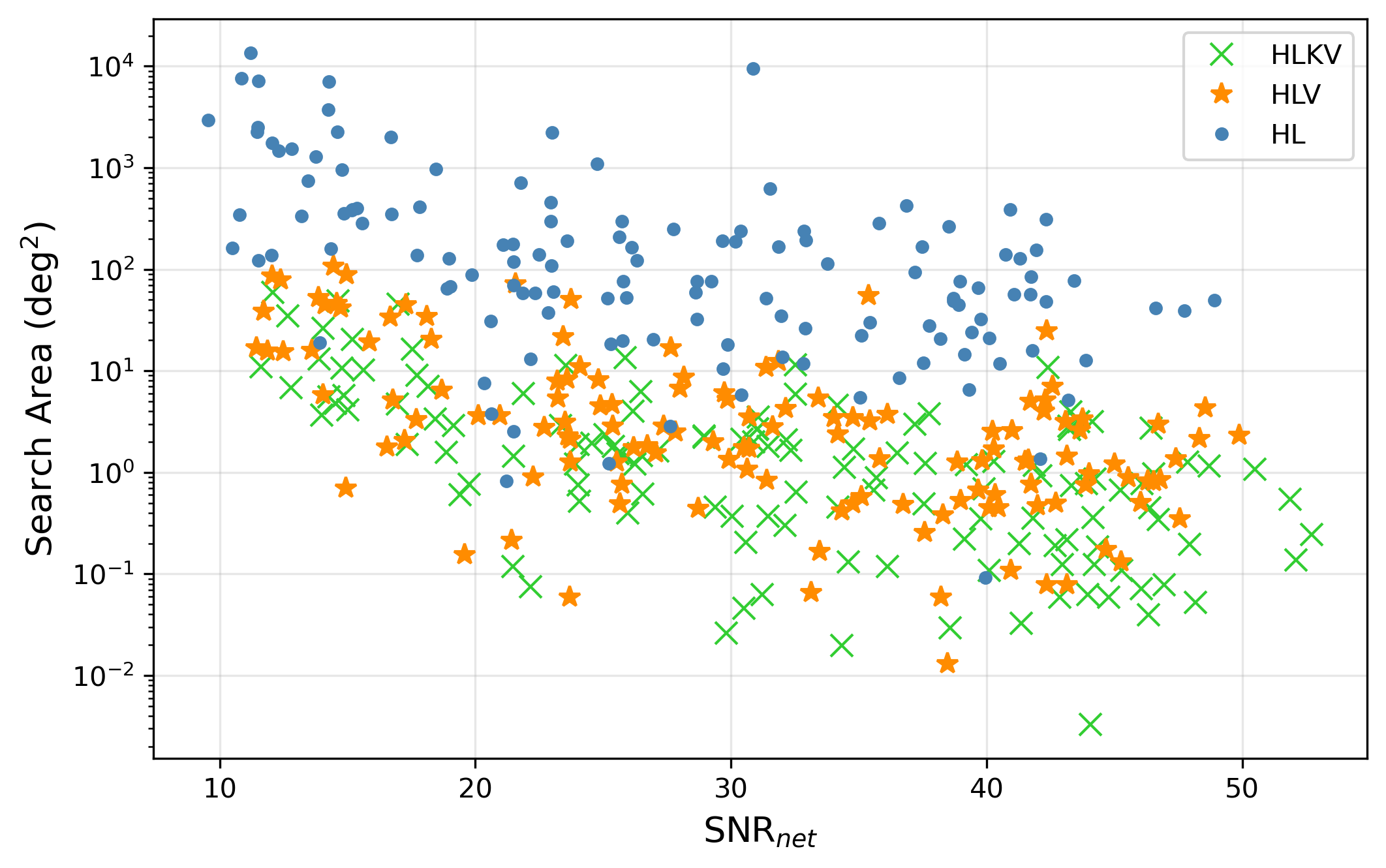}
    \caption{Search area, $\mathcal{A}$ (Equation \ref{eq:search_area}) versus network SNR for the HL (blue dots) and HLV (orange stars) networks.}
    \label{fig:searchArea}
\end{figure}

\subsection{Sky localisation} \label{sec:skyloc}

Expanding detector networks improves sky localisation of GW events, as has been shown by various studies on coherent network detections e.g \cite{2011CQGra..28l5023S}  \cite{2018ApJ...854L..25P} and \cite{2019PhRvD.100l4003P}; see Section \ref{sec:network}. In this section, we compare the accuracy of \textit{BayesWave} in locating the source with the HL and HLV networks. We use two separate measures: (i) sky area enclosed within the $50\%$ and $90\%$ credible intervals (CI) and (ii) search area, $\mathcal{A}$. 

For every injection, \textit{BayesWave} produces posterior distributions for the sky location (in the form of right ascension and declination) of the GW signal. We first look at the sky area enclosed within $50\%$ and $90\%$ credible intervals (CIs) of the posterior distribution of source location.
In the left panel of Figure \ref{fig:skyloc}, we show the plot for sky area enclosed within the $50\%$ CI versus network SNR for each injection, and similarly for the the  $90\%$ CI on the right panel. For all three detector configurations, we note that the area within the $50\%$ and $90\%$ CIs measured in square degrees (deg$^2$) fundamentally reduces with increasing network SNR due to improved accuracy in arrival time differences \cite{2018ApJ...854L..25P}. However, both sky areas are generally an order of magnitude smaller for the HLV network compared to the HL network. Upon addition of the KAGRA detector, we observe further reduction in the sky area, but not as drastic as that between the HL and HLV networks since KAGRA is less sensitive than Virgo. The areas enclosed within both $50\%$ and $90\%$ CIs reduces with increasing $\mathcal{I}$ due to the additional arrival time differences which further constrain the location of each source. These results reiterate that accuracy of sky localisation improves at fixed CI as $\mathcal{I}$ increases.

We also compare the inferred sky location with the true injected location of the source. We introduce another metric - the search area, $\mathcal{A}$, the hypothetical sky area observed by a detector before it correctly points towards the true location. To define this quantity mathematically, we first denote the posterior probability density function (PDF) of sky location as $p_{\text{sky}}(\phi, \theta)$. If the true location of the source is $(\phi_t, \theta_t)$ and  $p_0=p_{\text{sky}}(\phi_t, \theta_t)$, then all points within $\mathcal{A}$ should have $p_{\text{sky}}\geq p_0$. Mathematically, we write \cite{2015ApJ...800...81E, 2017ApJ...839...15B}
\begin{equation}
    \mathcal{A} = \int H[p_{\text{sky}}(\phi, \theta) - p_0] \, d\Omega
    \label{eq:search_area}
\end{equation}
where $H$ is the Heaviside step function and  $d\Omega$ is the surface area element on the celestial sphere i.e. $d\Omega = \cos\delta d\theta d\phi$. In Figure \ref{fig:searchArea} we plot the search area, $\mathcal{A}$ against network SNR for both the HL and HLV networks. The HLVK search area is slightly smaller than the HLV search area, which in turn is significantly smaller than the HL search area, consistent with Figure \ref{fig:skyloc}. 

Overall, we see that sky localisation improves remarkably when a detector of high-sensitivity is added to the network. If a less sensitive detector is added, the improvements are small but not negligible.

\section{Conclusion} \label{sec:summary}

The aim of this study is to compare the performance of \textit{BayesWave} in recovering GW waveforms from detector networks of different sizes. We derive an analytic scaling for the Bayes factor between the signal and glitch models, $\mathcal{B}_{\mathcal{S}, \mathcal{G}}$.  We then inject a set of simulated BBH signals of fixed masses at different SNRs into simulated O4 detector data of the HL, HLV and HLKV network. We quantify \textit{BayesWave}'s performance in signal identification with $\mathcal{B}_{\mathcal{S}, \mathcal{G}}$ and the performance in waveform reconstruction with overlap, $\mathcal{O}_\text{net}$. We also compare the accuracy of sky localisation between the two networks.

We find that events of similar injected SNR analysed using the HLV and HLKV network have higher $\ln \mathcal{B}_{\mathcal{S}, \mathcal{G}}$ than those using the HL network. This agrees with theoretical prediction of the Bayes factor scaling:
\begin{equation}
    \ln \mathcal{B}_{\mathcal{S},\mathcal{G}} \propto \mathcal{I}N\ln\text{SNR}_\text{net}.
    \label{eq:BFscaleapprox}
\end{equation}
Previous work~\cite{2016PhRvD..94d4050L} demonstrated that \textit{BayesWave} is unique amongst GW umodelled burst searches in that the so-called ``complexity'' of the signal in time-frequency plane plays a crucial role in the detection statistic, rather than just the signal's strength.  This is understood through the factor of $N$ in Equation~\ref{eq:BFscaleapprox}: a signal with more complex structure needs more wavelets to accurately reconstruct the waveform.  In this work, we expose another novel feature of the \textit{BayesWave} algorithm: the detection statistic is also influenced by the number of detectors i.e. the factor of $\mathcal{I}$ in Equation~\ref{eq:BFscaleapprox}.  Events of similar injected SNR (SNR$_\text{net}$) analysed using larger detector networks have higher $\ln \mathcal{B}_{\mathcal{S}, \mathcal{G}}$, indicating detection confidence increases more than we would expect purely from the increase in SNR$_\text{net}$.

The network overlap, $\mathcal{O}_\text{net}$, between the injected and recovered waveforms increases with SNR$_{\text{net}}$. We also show that $87\%$ of the HLKV network, $86\%$ of the HLV network and $81\%$ of the HL network injections have $\mathcal{O}>0.8$. Since larger detector networks can detect signals at higher SNR, they pick up more details of the true waveform. Thus, \textit{BayesWave} can reconstruct the waveforms more accurately.

Finally in Section \ref{sec:skyloc}, we quantify accuracy of sky localisation with the sky area enclosed within the $50\%$ and $90\%$ credible intervals (CI). We find that both areas decrease with increasing SNR$_\text{net}$ and are generally an order of magnitude smaller for the HLV networks than the HL network. The reduction of sky area is less significant upon the addition of the KAGRA detector due to its low sensitivity compared to Virgo. The search area, $\mathcal{A}$ also decreases with increasing SNR$_\text{net}$ and increasing number of detectors. The overall results suggest that increasing the number of detectors at different geographical locations improves sky localisation, consistent with previous analyses \cite{2011CQGra..28l5023S,2018ApJ...854L..25P, 2019PhRvD.100l4003P}.

With the global detector network growing in size, the outlook for improving detection confidence with unmodelled burst searches is promising. Prospective work along the lines of the research presented in this paper may include injecting different waveform morphologies to compare detection confidence between detector networks of different sizes. We also recommend looking into quantifying and comparing the outcomes of \textit{BayesWave} in recovering simulated signals from more realistic detector noise (i.e. in the presence of glitches) between different detector configurations.

In summary, \textit{BayesWave} shows significant improvements in terms of waveform recovery and parameter estimation when working with a larger detector network. This promising result suggests that with more detectors joining the global network in the future, we will be able to reconstruct generic GW burst signals more accurately using \textit{BayesWave} making detections with higher Bayes factor and hence with higher confidence.
\section*{Acknowledgements} \label{sec:acknowledgements}
    Parts of this research were conducted by the Australian Research Council Centre of Excellence for Gravitational Wave Discovery (OzGrav), through project number CE170100004.  The authors are grateful for computational resources provided by the LIGO Laboratory and supported by National Science Foundation Grants PHY-0757058 and PHY-0823459. We thank Bence B\'{e}csy for his helpful comments.

\appendix

\section{Fisher Information Matrix} \label{app:FIM}

Each wavelet has its Fisher Information Matrices (FIMs), $\Gamma$ written in terms of its five intrinsic parameters $\{t_0, f_0, Q, \ln A, \phi_0\}$
\begin{equation}
\Gamma = \text{SNR}^2
\begin{pmatrix}
\frac{4\pi^2f_0^2(1+Q^2)}{Q^2} & 0 & 0 &0 & -2\pi f_0\\
0 & \frac{3+Q^2}{4f_0^2} & -\frac{3}{4Qf_0} & -\frac{1}{2f_0} & 0\\
0 & -\frac{3}{4Qf_0} & \frac{3}{4Q^2} & \frac{1}{2Q} & 0 \\
0 & -\frac{1}{2f_0} & \frac{1}{2Q} & 1 & 0\\
-2\pi f_0 & 0 & 0 & 0 & 1
\end{pmatrix}_.
\end{equation}
FIMs contain information on local curvature of the likelihood of wavelet parameters which accelerates convergence by proposing jumps in the MCMC algorithm towards regions of higher likelihood \cite{2015CQGra..32m5012C}. \textit{BayesWave} uses FIMs to update wavelet parameters by drawing proposals from a multivariate Gaussian distribution 
\begin{equation}
    q(\mathbf{y}|\mathbf{x}) = \frac{\det \Gamma}{(2\pi)^2}\exp\left(-\frac{1}{2}\Gamma_{ij}\Delta x^i \Delta x^j\right)
\end{equation}
where $\Delta x^i = x^i - y^i$ denotes the displacement
in intrinsic parameter $i$ before and after the update. 

\section{Assumptions for Bayes Factor Scaling} \label{app:BF_scaling}

Laplace approximations for the logarithmic signal ($\mathcal{S}$) and glitch ($\mathcal{G}$) model evidences are given by Equations \ref{eq:evidence_s} and \ref{eq:evidence_g} respectively. In order to see how $\mathcal{B}_{\mathcal{S}, \mathcal{G}}$ scales with the waveform parameters, we make some assumptions to simplify the two logarithmic evidences. In this work we use the same assumptions as in Ref. \cite{2016PhRvD..94d4050L}. 

Loud signals typically have optimal extrinsic parameters across the detector network, so the SNR in each detector will be approximately equal such that
\begin{equation}
    \text{SNR}_{i,n} \approx \frac{\text{SNR}_{\text{net},n}}{\sqrt{\mathcal{I}}}
    \label{eq:ifo_wavelet_snr}
\end{equation}
where $\text{SNR}_{i,n}$ is the SNR of the $n$-th wavelet in detector $i$.
We use a further simplifying assumption that the SNR of each wavelet is the same 
\begin{equation}
    \text{SNR}_{\text{net},n} \approx  \frac{\text{SNR}_\text{net}}{\sqrt{N}},
\end{equation}
which has been empirically validated.
We assume that the glitch model in each detector uses similar reconstruction parameters as the signal model, and as such the quality factors of all wavelets are approximately equal:
\begin{equation}
    Q^\mathcal{G}_{i,n} \approx Q^\mathcal{S}_n \equiv Q
\end{equation}
and similarly,  
\begin{equation}
    N^\mathcal{G} \approx  N^\mathcal{S} \equiv N.
    \label{eq:Ng_Ns}
\end{equation}
Recall that $N^\mathcal{G}$ indicates the number of wavelets used in the glitch model for a \textit{single} detector, so for an $\mathcal{I}$-detector network, the total number of wavelets used in glitch models across the entire network is $\mathcal{I} N$.

Equations \ref{eq:evidence_s} and \ref{eq:evidence_g} can be simplified to
\begin{widetext}
\begin{equation}
    \ln p\inprod{d}{\mathcal{S}} \simeq  \frac{\text{SNR}_\text{net}^2}{2} - \frac{5N}{2} - N\ln(V_\lambda) + \sum _{n=1} ^{N} \ln \left( \bar{Q}_n\right) - 5N\ln (\frac{\text{SNR}_\text{net}}{\sqrt{N}})+ \left(2 + \ln \frac{\sqrt{\det C_\Omega}}{V_\Omega}\right)
\label{eq:evidence_s_cut}
\end{equation}

\begin{equation}
\ln p\inprod{d}{\mathcal{G}} \simeq  \frac{\text{SNR}_\text{net}^2}{2} - \mathcal{I} \left[\frac{5N}{2} + N\ln(V_\lambda) - \sum _{n=1}  ^{N} \ln \left( \bar{Q}_n\right) + 5N\ln (\frac{\text{SNR}_\text{net}}{\sqrt{N\mathcal{I}}})\right].
\label{eq:evidence_g_cut}
\end{equation}
\end{widetext}

\section{Scaling of Bayes factor with $\mathcal{I}$} \label{app:SG_injections}

\begin{figure}[h]
\centering
\includegraphics[width=0.49\textwidth]{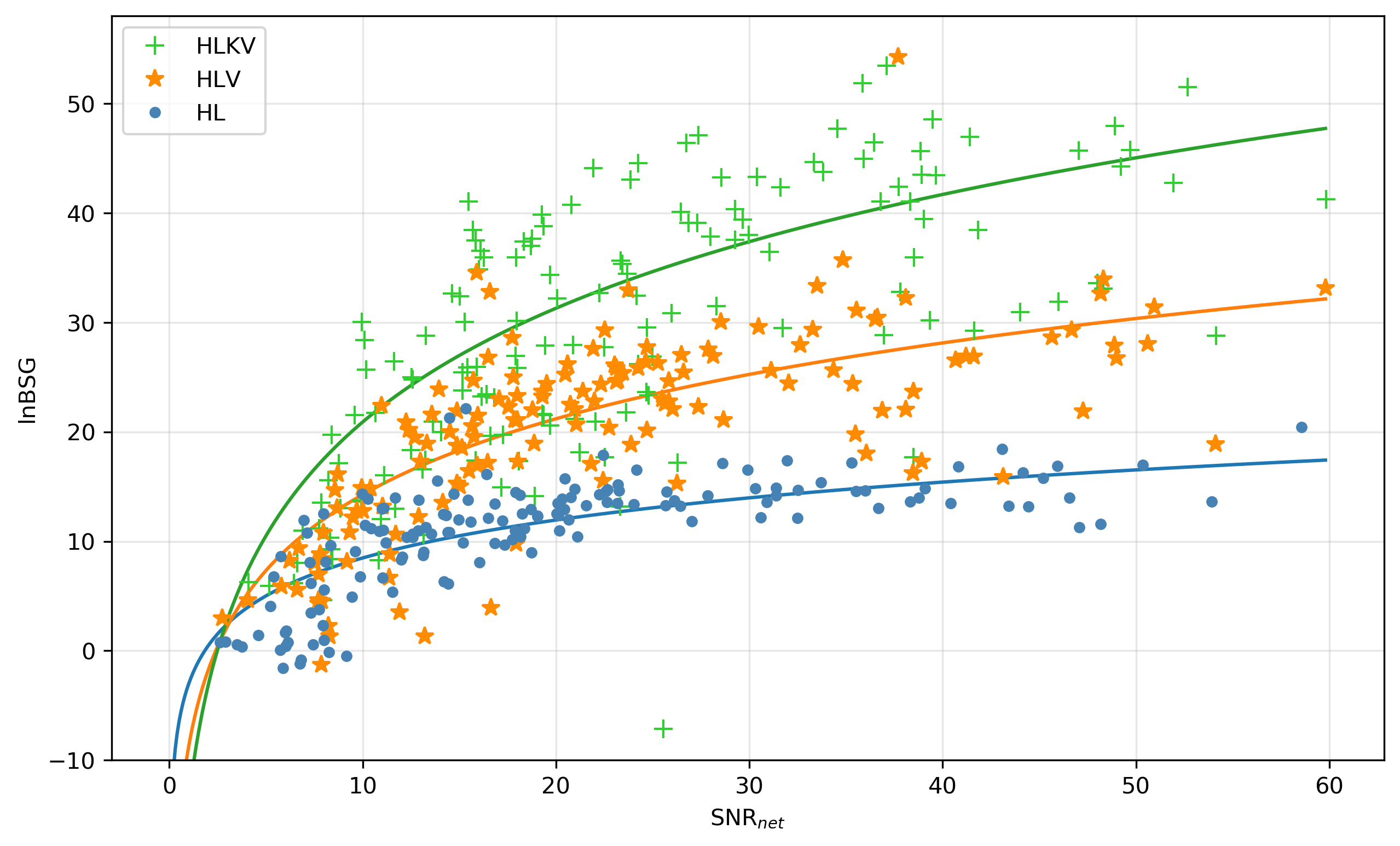}
\caption{Log signal-to-glitch Bayes factor, $\ln \mathcal{B}_{\mathcal{S},\mathcal{G}}$ of sine-Gaussian wavelet recoveries versus network signal-to-noise ratio, SNR$_{\text{net}}$. The solid lines with colours corresponding to the data symbols are analytic predictions of $\ln \mathcal{B}_{\mathcal{S},\mathcal{G}}$ given by Equation \ref{eq:appx_BFscaling}.}
\label{fig:SG_injections}
\end{figure}

Our results in Section \ref{sec:BayesFactor} show that $\ln \mathcal{B}_{\mathcal{S},\mathcal{G}}$ scales with SNR, $N$, and $\mathcal{I}$. As per Equation~\ref{eq:wavelet_complexity}, $N$ itself depends on both the SNR of the signal, and the waveform morphology. In order to specifically test the scaling of $\ln \mathcal{B}_{\mathcal{S},\mathcal{G}}$ with $\mathcal{I}$ alone, we inject a set of sine-Gaussian wavelets
as coherent signals into detector noise for the HL, HLV and HLKV network and then recover them using \textit{BayesWave}. Because sine-Gaussian wavelets are the basis of reconstruction for \textit{BayesWave}, the number of wavelets used is $N=1$, with no dependence on SNR.

The dataset used this analysis is a set of 150 single sine-Gaussian wavelets. The parameters of each wavelet are randomly drawn from the following distributions: $t_0 \in [1.5, 2.5]$ s (where $t=1\ {\rm s}$ is the center of the analysis window), $f_0 \in [32, 1000]$ Hz, $Q \in [0.1, 40]$ and $\phi_0 \in [0, 2\pi]$. The SNR of the signals are drawn randomly from a uniform distribution and SNR $\in [10,50]$, and the amplitude is then found viz. 
\begin{equation}
A = \text{SNR}\sqrt{\frac{2\sqrt{2\pi}f_0S_n(f_0)}{Q}}
\label{eq:amplitude_prior}
\end{equation} 
(see \cite{2018PhRvD..97j4057M} for details). As we are injecting a coherent signal, we also require four extrinsic parameters as described in Section~\ref{subsec:Wavelets}. These parameters are also drawn randomly from uniform distributions such that $\alpha \in[0,2\pi]$, $\delta \in [-\sfrac{\pi}{2}, \sfrac{\pi}{2}]$, $\psi \in [0, 2\pi]$ and $\epsilon \in [-0.99, 0.99]$. 

In Figure~\ref{fig:SG_injections}, we plot 
$\ln \mathcal{B}_{\mathcal{S},\mathcal{G}}$ 
of each injection against ${\rm SNR}_\text{net}$
for the HL, HLV and HLKV network injections.  We note that $\ln \mathcal{B}_{\mathcal{S},\mathcal{G}}$ increases with SNR$_\text{net}$ and is generally higher for networks with greater $\mathcal{I}$, as predicted from Equation~\ref{eq:BF_scaling2}. 
Since $N$ in this case does not depend on SNR, we can be certain that the differences in $\ln \mathcal{B}_{\mathcal{S},\mathcal{G}}$ between the different detector configurations at comparable SNR$_\text{net}$ are entirely due to $\mathcal{I}$. 

In order to compare the analytic and empirical scaling of $\ln \mathcal{B}_{\mathcal{S},\mathcal{G}}$ with $\mathcal{I}$, we fit analytic approximation of $\ln \mathcal{B}_{\mathcal{S},\mathcal{G}}$ for each detector network with a generalised expression
\begin{equation}
\ln \mathcal{B}_{\mathcal{S},\mathcal{G}} \approx (\mathcal{I}-1)[5\ln\text{SNR}_\text{net} + a] + \frac{5}{2}\mathcal{I}\ln\mathcal{I} + b.
\label{eq:appx_BFscaling}
\end{equation}
This expression is derived from Equation \ref{eq:BF_scaling} with $N=1$. 
We define the constants
$a = \frac{5}{2} - \ln(V_\lambda) + \ln(Q)$
and $b=2 + \ln \frac{\sqrt{\det C_\Omega}}{V_\Omega}$.  The prior volumes, $V_\lambda$ and $V_\Omega$ are respectively the same for all detector configurations.  We do not have an analytic expression for $\det C_\Omega$ as the FIM approximations is inadequate; the extrinsic parameter space contains degeneracy between parameters, resulting in multimodal, non-Gaussian likelihood distribution which spans the full extent of the prior range (see \cite{2016PhRvD..94d4050L} for details).
We present the fits as three solid lines in Figure \ref{fig:SG_injections}, where we have determined by-eye $a=-10$ and $b=4$.
The same values of $a$ and $b$ are used for all three fits and they 
are broadly consistent with the empirical results.

We see general agreement between the empirical results and predicted scaling for $\ln \mathcal{B}_{\mathcal{S},\mathcal{G}}$, which further confirms our results in Section~\ref{sec:BayesFactor} that the Bayes factor does not only depend on SNR$_\text{net}$ but also on $\mathcal{I}$.
We again note that these scalings are estimations, and due to the different sensitivities of the detectors we do not expect exact agreement between analytic prediction and empirical results.

\bibliographystyle{unsrt}
\bibliography{citations}

\end{document}